\documentclass[12pt]{iopart}
\usepackage{amsfonts}
\usepackage{amssymb}
\usepackage{graphicx}
\usepackage{bm}
\usepackage{color}

\begin{document}

\title{Quantum Rabi-Stark model: Solutions and exotic energy spectra}
\author{You-Fei  Xie$^{1}$, Liwei Duan$^{1,2}$, and Qing-Hu Chen$^{1,3,*}$}

\address{$^1$ Department of Physics and Zhejiang Province Key Laboratory of Quantum Technology and Device, Zhejiang University, Hangzhou 310027, China }
\address{$^{2}$ Department of Physics, Zhejiang Normal University, Jinhua 321004, China}
\address{$^3\emph{}$ Collaborative Innovation Center of Advanced Microstructures,  Nanjing University,  Nanjing 210093, China}
\eads{\mailto{qhchen@zju.edu.cn}}
\date{\today }

\begin{abstract}
The quantum Rabi-Stark model, where the linear dipole coupling and the
nonlinear Stark-like coupling are present on an equal footing, are studied
within the Bogoliubov operators approach. Transcendental functions
responsible for the exact solutions are derived in a compact way, much
simpler than previous ones obtained in the Bargmann representation. The
zeros of transcendental functions reproduce completely the regular spectra.
In terms of the explicit pole structure of these functions, two kinds of
exceptional eigenvalues are obtained and distinguished in a transparent
manner. Very interestingly, a  first-order quantum phase transition indicated
by level crossing of the ground state and the first excited state is induced
by the positive nonlinear Stark-like coupling, which is however absent in any previous isotropic quantum Rabi models. When the absolute value of
the nonlinear coupling strength is equal to  twice the cavity frequency,
this model can be  reduced to an effective quantum harmonic
oscillator, and solutions are then obtained analytically. The spectra collapse phenomenon is observed at a  critical  coupling, while below this  critical coupling, infinite discrete spectra accumulate into a finite energy  from below.
\end{abstract}


\vspace{2pc}

\maketitle

\section{Introduction}

The quantum Rabi model (QRM), which represents the simplest interaction
between a two-level atom (qubit) and a light field (cavity), continues to
inspire exciting developments in many fields ranging from quantum optics,
quantum information science, and condensed matter physics ~\cite{Braak2}.
The Hamiltonian is given by
\begin{equation}
H_{R}=\frac{\Delta }{2}\sigma _{z}+\omega a^{\dagger }a+g\left( a^{\dagger
}+a\right) \sigma _{x},  \label{QRM}
\end{equation}%
where $\Delta $ $\ $\ and $\omega \ $\ are frequencies of two-level system
and cavity, $\sigma _{x,z}$ are usual Pauli matrices describing the
two-level system, $a$ ($a^{\dagger }$) is the annihilation (creation)
bosonic operator of the cavity mode, and $g$ is the coupling strength. In
the conventional cavity quantum electrodynamics (QED) system \cite{CQED},
the coupling strength between the atom and the field is quite weak, $%
g/\omega \thicksim 10^{-6}$. It can be described by the well-known
Jaynes-Cummings model~\cite{JC} where the rotating-wave approximation is
made. Many physical phenomena can be described in this framework, such as
collapse and revival of quantum state populations, vacuum Rabi splitting,
and photon anti-bunching~\cite{Scully}.

With the progress of the experimental techniques, the QRM can be implemented
in enhanced parameter regimes. Some solid-state devices such as
superconducting circuits \cite{Niemczyk,exp,Yoshihara,Forn2,tiefu}, quantum
wells \cite{Anappara}, cold atoms \cite{cold} have emerged as genuine
platforms for faithful representations of this model in the ultra-strong $%
\left( g/\omega \thicksim 0.1\right) $, even deep-strong-coupling $\left(
g/\omega >1\right) $ regime \cite{Casanova}. Evidence for the breakdown of
the rotating-wave approximation has been provided in the qubit-oscillator
system at ultra-strong coupling \cite{Niemczyk}. Many works then have been
devoted to this system in the ultra-strong coupling regime ~\cite%
{Werlang,Hanggi,Ashhab,Hausinger,chen10b}. Recently, the competition to
increase the coupling strength is still on-going in different experimental
systems ~\cite{Yoshihara,Forn2,Bayer,Nori}.

On the other hand, quantum simulations can engineer the interactions in a
well-defined quantum system to implement the target model of interest in the
infeasible parameter regime \cite{Puebla}. The engineered system even
enables the generalization of the target model, thus more fundamental
phenomena might emerge. The QRM with arbitrary parameters has been realised
in quantum simulations based on Raman transitions in an optical cavity QED
settings~\cite{Grimsmo1,Grimsmo}. In this proposed scheme ~\cite{Grimsmo1},
beside the linear dipole coupling, the following nonlinear coupling between
atom and field can also emerge
\begin{equation}
H_{NL}=\frac{U}{2}\sigma _{z}a^{\dagger }a,  \label{NL}
\end{equation}%
where the coupling strength $U$ is determined by the dispersive energy
shift. It is associated with the dynamical Stark shift discussed in the
quantum optics \cite{Klimov}, so this generalized model proposed by Grimsmo
and Parkins is called quantum Rabi-Stark model \cite{Eckle}. This emergent
Stark-like nonlinear interaction has no parallel in the conventional cavity
QED, which adds a new member to the list of various quantum Rabi models.

Any modification to the linear QRM described by Hamiltonian (\ref{QRM})
would possibly bring about the novel and exotic physical properties. The
interaction-induced energy spectral collapse can be observed in the
two-photon QRM when the normalized coupling approaches the half of cavity
frequency~\cite{Ng,Felicetti,duan}. The anisotropic QRM, where the coupling
strength of the rotating-wave terms and counter-rotating wave terms is
different, exhibits the first-order phase transitions ~\cite{Fanheng}. These
phenomena are obviously absent in the original linear isotropic QRM ~\cite%
{Braak2}. Grimsmo and Parkins conjecture that the nonlinear coupling
manipulated by the dispersive energy shift would possibly induce a new
superradiant phase at this single atom level if $U<-2\omega $~\cite{Grimsmo1}%
. Although the total Hamiltonian $H_{0}=H_{R}+H_{NL}$ has been studied by
the Bargmann approach~\cite{Eckle,Maciejewski21}, no much attention has yet
been paid to its possible novel and peculiar physical properties, to the
best of our knowledge.

Analytical solutions to the linear QRM have been searched for a few decades
(for a review, please refer to Refs. ~\cite{Lee,yuxi,ReviewF}). Many
approximate analytical solutions have been proposed \cite%
{Swain,Kus,Durstt,Feranchuk,Bishop,Irish,Paganelli,chen10,Pan,chen11,luo2,zhang}%
, however analytically exact solution was only found by Braak ~\cite%
{Braak2011} using the Bargmann representations. It was shown that Braak's
solution can be constructed in terms of the mathematically well-defined Heun
confluent function~\cite{Zhong}. By Bogoliubov operators approach (BOA) ~%
\cite{Chen2012}, Braak's solution was reproduced straightforwardly in a more
transparent manner. One clear advantage is that for a discussion of the BOA,
it is not required to refer to heavy mathematical terminology. It is
generally accepted in the literature that the BOA is more physical ~\cite%
{Braak2, Villas}. Moreover, BOA can be easily extended to the two-photon QRM~%
\cite{Chen2012}, and solutions in terms of a $G$-function, which shares the
common pole structure with Braak's $G$-function for the one-photon QRM, are
also found. It was demonstrated later in ~\cite{Felicetti} that this
two-photon $G$-function by BOA ~\cite{Chen2012} allows for the desired
understanding of the qualitative features of the collapse. However, the $G$%
-function by the direct application of the Bargmann space approach ~\cite%
{Trav} has no pole structure, and thus could not give qualitative insight
into the behavior of the spectral collapse ~\cite{Felicetti}. To the best of
our knowledge, the G-function with its pole structure for the two-photon QRM
has only been found using the BOA and, in particular, has so far not been
derived using the Bargmann space method. So in the study of the anisotropic
two-photon QRM, only the BOA is employed ~\cite{Cui}. In principle, the
Bargmann space approach could still be used to recover the correct
G-function in the two-photon QRM, which might require more mathematics.

In this work, we will study the quantum Rabi-Stark model by the BOA, and
then explore some exotic physical phenomena. The paper is structured as
follows: In section II, a concise $G$-function is derived for this model by
using BOA. In section III, two kinds of exceptional solutions are obtained
explicitly in terms of the pole structure of the obtained transcendental
function. First-order phase transitions are then analytically detected. The
energy spectral collapse is discussed by an effective one-body Hamiltonian
corresponding to a quantum harmonic oscillator in Sec. IV. The last section
contains some concluding remarks and outlooks.

\section{Bogoliubov operators approach and $G$-function}

To facilitate the study, the Hamiltonian $H_0=H_{R}+H_{NL}$ is rotated
around the y-axis by an angle $\pi /2$
\begin{equation}
H=-\frac{1}{2}\left( \Delta +Ua^{\dagger }a\right) \sigma _{x}+\omega
a^{\dagger }a+g\left( a^{\dagger }+a\right) \sigma _{z}.  \label{Stark-Rabi}
\end{equation}%
In terms of two eigenstates of $\sigma _{z}$, the above Hamiltonian takes
the following matrix form in units of$\ \omega =1$
\begin{equation}
H=\left(
\begin{array}{ll}
a^{\dagger }a+g\left( a^{\dagger }+a\right) & ~-\frac{1}{2}\left( \Delta
+Ua^{\dagger }a\right) \\
-\frac{1}{2}\left( \Delta +Ua^{\dagger }a\right) & a^{\dagger }a-g\left(
a^{\dagger }+a\right)%
\end{array}%
\right) .  \label{matrix}
\end{equation}%
Associated with this Hamiltonian is the conserved parity $\Pi =\exp \left(
i\pi \widehat{N}\right) \ $where $\widehat{N}=\left( 1-\sigma _{x}\right)
/2+a^{\dagger }a$ is the total excitation number, such that $\left[ \Pi ,H%
\right] =0$. $\Pi $ has two eigenvalues $\pm 1$, depending on whether $%
\widehat{N}$ is even or odd.

We first perform the Bogoliubov transformation with displacement $w$%
\begin{equation}
A=a+w,  \label{A}
\end{equation}%
where $A$ is the new bosonic operator which obeys the commutation relation $%
\left[ A,A^{\dag }\right] =1$, the shift $w$ will be determined later. The
transformed Hamiltonian then reads
\begin{equation}
H=\left(
\begin{array}{ll}
H_{11} & H_{12} \\
H_{21} & H_{22}%
\end{array}%
\right) ,
\end{equation}%
where
\begin{eqnarray*}
H_{11} &=&A^{\dagger }A+\left( g-w\right) \left( A^{\dagger }+A\right)
+w^{2}-2gw, \\
H_{12} &=&H_{21}=-\frac{\Delta }{2}-\frac{U}{2}\left[ A^{\dagger }A-w\left(
A^{\dagger }+A\right) +w^{2}\right] ,\  \\
H_{22} &=&A^{\dagger }A-\left( g+w\right) \left( A^{\dagger }+A\right)
+w^{2}+2gw.
\end{eqnarray*}%
The wavefunction can be expanded in terms of the $A$-operators
\begin{equation}
\left\vert {}\right\rangle _{A}=\left( \
\begin{array}{l}
\sum_{n=0}^{\infty }\sqrt{n!}e_{n}\left\vert n\right\rangle _{A} \\
\sum_{n=0}^{\infty }\sqrt{n!}f_{n}\left\vert n\right\rangle _{A}%
\end{array}%
\right) ,  \label{wave1}
\end{equation}%
where $e_{n}$ and $f_{n}$ are the expansion coefficients. $\left\vert
n\right\rangle _{A}$ is called extended coherent state \cite{chenqh2} with
the following properties
\begin{eqnarray}
\left\vert n\right\rangle _{A} &=&\frac{\left( a^{\dagger }+w\right) ^{n}}{%
\sqrt{n!}}\left\vert 0\right\rangle _{A},  \label{ex2} \\
\left\vert 0\right\rangle _{A} &=&e^{-\frac{1}{2}w^{2}-wa^{\dagger
}}\left\vert 0\right\rangle _{a},  \nonumber
\end{eqnarray}%
where the vacuum state $\left\vert 0\right\rangle _{A}$ in Bogoliubov
operators $A$ is just well-defined as the eigenstate of one-photon
annihilation operator $a$, and known as pure coherent state.

Projecting both sides of the Schr\"{o}dinger equation onto $_{A}\left\langle
m\right\vert \ $ gives%
\begin{equation}
\left( \Gamma _{m}-E-2gw\right) e_{m}+\left( g-w\right) \Lambda _{m}-\left(
\frac{\Delta }{2}+\frac{U}{2}\Gamma _{m}\right) f_{m}+\frac{U}{2}w\digamma
_{m}=0,  \label{S1}
\end{equation}%
\begin{equation}
-\left( \frac{\Delta }{2}+\frac{U}{2}\Gamma _{m}\right) e_{m}+\frac{U}{2}%
w\Lambda _{m}+\left( \Gamma _{m}-E+2gw\right) f_{m}-\left( g+w\right)
\digamma _{m}=0,  \label{S2}
\end{equation}%
where
\begin{eqnarray*}
\Lambda_{m}&=&(m+1)e_{m+1}+e_{m-1}, \\
\quad \digamma _{m} &=&(m+1)f_{m+1}+f_{m-1}, \\
\quad \Gamma _{m} &=&m+w^{2}.
\end{eqnarray*}%
To get one-to-one correspondence of $e_{m}$ and $f_{m}$, one should cancel
the terms involving $\Lambda _{m}$ and $\digamma _{m}$, which requires the
shift $w$ to be%
\begin{equation}
w=\frac{g}{\sqrt{1-U^{2}/4}}.  \label{shift}
\end{equation}%
It is just equal to the value of the singularity in \cite{Eckle}. Then we
have
\begin{equation}
e_{m}=\Omega _{m}f_{m},  \label{enfn}
\end{equation}%
where%
\begin{equation}
\Omega _{m}=\frac{\frac{Uw}{g+w}\left( \Gamma _{m}-E+2gw\right) -\left(
\Delta +U\Gamma _{m}\right) }{\frac{Uw}{2\left( g+w\right) }\left( \Delta
+U\Gamma _{m}\right) -2\left( \Gamma _{m}-E-2gw\right) }.  \label{omega}
\end{equation}%
Inserting Eq. (\ref{enfn}) \ into Eq. (\ref{S1}), we obtained a three-term
recurrence relation for $f_{m}$
\begin{eqnarray}
f_{m} &=&\frac{\Delta +U\Gamma _{m-1}-2\left( \Gamma _{m-1}-E-2gw\right)
\Omega _{m-1}}{m\left[ Uw+2\left( g-w\right) \Omega _{m}\right] }f_{m-1}
\nonumber \\
&&-\frac{2\left( g-w\right) \Omega _{m-2}+Uw}{m\left[ Uw+2\left( g-w\right)
\Omega _{m}\right] }f_{m-2}  \label{recur}
\end{eqnarray}%
where all $f_{m}$ can be obtained if set $f_{0}=1$.

By the opposite shift $(-w)$, we can define another Bogoliubov operator
\begin{equation}
B=a-w,  \label{B}
\end{equation}%
the wavefunction can also be expanded in the $B$-basis as
\begin{equation}
\left\vert {}\right\rangle _{B}=\left( \
\begin{array}{l}
\sum_{n=0}^{\infty }(-1)^{n}\sqrt{n!}f_{n}\left\vert n\right\rangle _{B} \\
\sum_{n=0}^{\infty }(-1)^{n}\sqrt{n!}e_{n}\left\vert n\right\rangle _{B}%
\end{array}%
\right) ,  \label{wave2}
\end{equation}%
due to the parity symmetry. $\left\vert n\right\rangle _{B}$ is defined
similar to $\left\vert n\right\rangle _{A}$.

Assuming both wavefunctions (\ref{wave1}) and (\ref{wave2}) are the true
eigenfunction for a nondegenerate eigenstate with eigenvalue $E$, they
should be proportional with each other, \textsl{i.e.} $\left\vert
{}\right\rangle _{A}=r\left\vert {}\right\rangle _{B}$, where $r$ is a
complex constant. Projecting both sides of this identity onto the original
vacuum state $_{a}\left\langle 0\right\vert $, we have
\begin{eqnarray*}
\sum_{n=0}^{\infty }\sqrt{n!}e_{n}~_{a}\langle 0|n\rangle _{A}
&=&r\sum_{n=0}^{\infty }\sqrt{n!}(-1)^{n}f_{n}~_{a}\langle 0|n\rangle _{B},
\\
\sum_{n=0}^{\infty }\sqrt{n!}f_{n}~_{a}\langle 0|n\rangle _{A}
&=&r\sum_{n=0}^{\infty }\sqrt{n!}(-1)^{n}e_{n}~_{a}\langle 0|n\rangle _{B},
\end{eqnarray*}%
where
\[
\sqrt{n!}~_{a}{\langle }0|n{\rangle }_{A}=(-1)^{n}\sqrt{n!}~_{a}{\langle }0|n%
{\rangle }_{B}=e^{-w^{2}/2}w^{n}.
\]
Eliminating the ratio constant $r$ gives
\[
\left( \sum_{n=0}^{\infty }e_{n}w^{n}\right) ^{2}=\left( \sum_{n=0}^{\infty
}f_{n}w^{n}\right) ^{2}.
\]%
Immediately, we obtain the following well-defined transcendental function,
so called $G$-function
\begin{equation}
G_{\mp }\left( E\right) =\sum_{n=0}^{\infty }\left( \Omega _{n}\pm 1
\right)f_{n} w^{n}=0,  \label{G-func}
\end{equation}%
where $\Omega _{n}$ and $f_{n}$ can be obtained from Eqs. (\ref{omega}) and (%
\ref{recur}), $\mp $ corresponds to negative(positive) parity. The zeros of
this $G$-function will give the regular spectrum, which should be the same
as those in ~\cite{Eckle}. In principle, there are many G-functions, all
have the same zeros and yield the same spectrum \cite{ann2013}. Note also
that this $G$-function can be reduced to that of the original QRM ~\cite%
{Braak2011} if set $U=0$.

\begin{figure}[tbp]
\centering
\includegraphics[width=8.5cm]{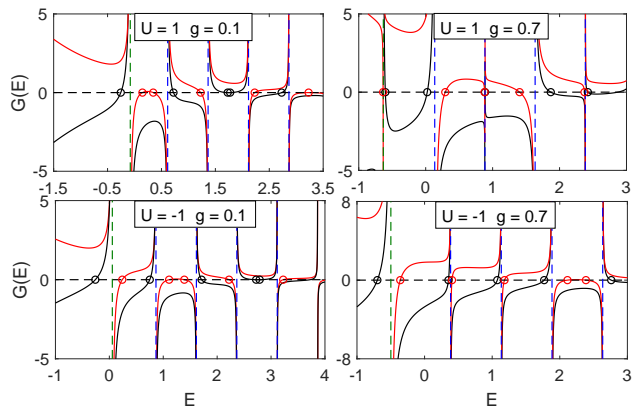}
\caption{ (Color online) G-curves for $\Delta =0.5$, $U =1$ (upper panels)
and $U=-1$ (lower panels), $g=0.1$ (left panels ) and $g=0.7$ (right
panels). Black lines and Red lines are $G_+$ and $G_-$ cures respectively.
The green dashed line is $E_{0}^{pole}$ and the blue dashed lines are $%
E_{n}^{pole}$. The data by numerics are indicated by circles, which agree
excellently with the zeros of the $G$-functions.}
\label{G-function}
\end{figure}

$G$-curves for $\Delta =0.5$, $U=\pm 1$, $g=0.1$ and $0.7$ are plotted in
Fig. \ref{G-function}. The zeros are easily detected, and then regular
energy spectra are obtained, which are exhibited in Fig. \ref{spectra}. As
usual, one can check it easily with numerics, an excellent agreement can be
achieved.

\section{ Exceptional solutions and first-order phase transitions}

From Eq. (\ref{shift}), we can note that the present solution by BOA can be
only applied to the Rabi-Stark model for $\left\vert U\right\vert <2$. Let
us now discuss novel features of the derived $G$-functions and the
exceptional spectra.

\subsection{Pole structure}

We first examine the pole structure of the  $G$-function (\ref{G-func}).
Note from Eq. (\ref{recur}) that the denominator of $f_{n}$ for $n>0$
vanishes, yielding the $n$-th pole of the $G$-function
\begin{equation}
E_{n}^{pole}=\left( 1-\frac{U^{2}}{4}\right) n-\frac{U\Delta }{4}-g^{2},
\label{pole}
\end{equation}%
It is interesting to find that this pole is reduced to $E_{n}^{QRM}=n-g^{2}$%
, the pole of the pure QRM ~\cite%
{Braak2011}, if set $U=0$.

From Eq. (\ref{omega}), one can find that $\Omega _{n}$ diverges at
\begin{equation}
E_{\Omega _{n}}=\frac{E_{n}^{pole}}{\sqrt{1-U^{2}/4}}+\frac{\Delta U}{%
4-U^{2}+4\sqrt{1-U^{2}/4}}.  \label{pole2}
\end{equation}%
However it is not the pole of the $G$-function, because $\Omega _{n}f_{n}$
appears always as a whole in the $G$-function (\ref{G-func}) and  is finite
at $E=E_{\Omega _{n}}$ for $n\neq 0$.

\begin{figure}[tbp]
\centering
\includegraphics[width=8.5cm]{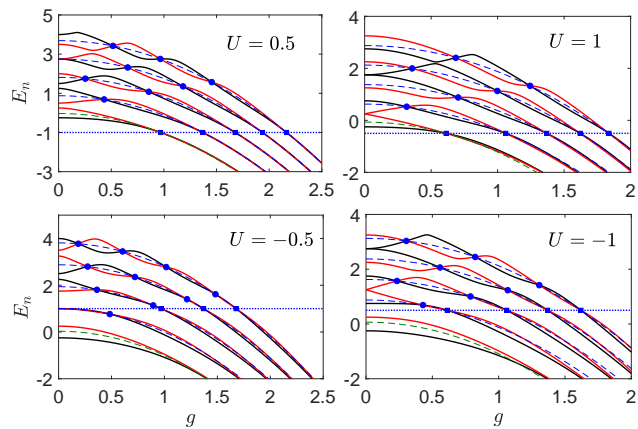}
\caption{ (Color online) Energy spectra for positive (upper panels) and
negative (lower panels) $U$ where $\Delta =0.5$. Red (negative) and black
(positive) denote different parity. The green dashed line is $E_{0}^{pole}$
and the blue dashed lines are $E_{n}^{pole}$ for $n=1,2,3,4 $. Level
crossings are marked by filled symbols. First-order phase transitions are
present (absent) for positive (negative) $U$. Horizontal blue dotted lines $%
E=-\Delta/U$ are guides to the eye. Exceptional solutions for nondegenerate
states are not given here. }
\label{spectra}
\end{figure}

Eq. (\ref{pole}) is not suited to $n=0$, because $f_{0}=1$. Particularly, the first term in the $G$%
-function (\ref{G-func}), $\Omega _{0}\pm 1$, really diverges at
\begin{equation}
E_{0}^{pole}=-\frac{g^{2}+\Delta U/4}{\sqrt{1-U^{2}/4}}+\frac{\Delta U}{%
4-U^{2}+4\sqrt{1-U^{2}/4}},  \label{0pole}
\end{equation}%
which is just the zeroth pole of the $G$-function.

The poles given in Eqs. (\ref{pole}) and (\ref{0pole}) are marked with
vertical lines in the $G$-curves of Fig. \ref{G-function}. The $G$-curves
indeed cannot pass through these poles, therefore the whole G-curves are
blocked into different smooth segments.

\subsection{Exceptional solutions}

\subsubsection{Juddian solutions for doubly degenerate states}

If the true physical system takes the energy at the zeroth pole $%
E_{0}^{pole} $, the wavefunction (\ref{wave1}) including $e_{0}=\Omega
_{0}f_{0}$ terms should be analytic. Hence both the denominator and
numerator of $\Omega _{0}$ should vanish at the same time, yielding the
constrained condition for the model parameter
\begin{equation}
g_{c}=\sqrt{\frac{\left( 1-U^{2}/4\right) }{U}\Delta }.  \label{gc}
\end{equation}%
Inserting Eq. (\ref{gc}) into Eq. (\ref{0pole}) gives the energy without specified parity $%
E_{0}^{cross}=-\frac{\Delta }{U}$. The first energy levels for both parities
thus intersect at $g_{c}$. These are the doubly degenerate states,
corresponding to the Juddian solution~\cite{Judd}.

Physically, the energies for the ground-state and the first excited state
cross, indicating a first-order quantum phase transition. According to Eq. (%
\ref{gc}), note that the qubit frequency $\Delta $ is always positive, so
the finite real $g_{c}$ only exists for $U>0$. No  first-order phase
transition exists in the present model for $U<0$ and the linear QRM
where $U=0$. As shown in the upper panels of Fig. \ref{spectra}, two levels
for the first excited state and ground state really cross once for $U>0$.
Such a crossing for the two lowest levels does not occur for $U<0$, as shown
in the lower panels.

Actually, for any $n$, if both denominator and numerator of $\Omega _{n}$ in
Eq. (\ref{omega}) vanish, $\Omega _{n}$ is analytic, leading to analytic
coefficients $e_{n}$ and $f_{n}$. The reason is the following. $\Omega
_{n}=x_{n}/y_{n}$ is analytic for both $x_{n}=0$ and $y_{n}=0$. The
denominator of $f_n$ in Eq. (\ref{recur}) is
\[
Uw+2\left( g-w\right) \Omega _{n}=\frac{Uwy_{n}+2\left( g-w\right) x_{n}}{%
y_{n}},
\]
we can easily find that both denominator and numerator of $f_{n}$ should be
also zero, leading to analytic coefficients $f_{n}$, and therefore analytic
coefficients $e_{n}$.

The condition that both the denominator and numerator of\ $\Omega _{n}$ in$\
$Eq. (\ref{omega}) vanish will give the coupling strength
\begin{equation}
g_{c}^{(n)}=\sqrt{\left( n+\frac{\Delta }{U}\right) \left( 1-\frac{U^{2}}{4}%
\right) }.  \label{gcn}
\end{equation}%
The corresponding energy $E_{n}^{cross}=-\frac{\Delta }{U}$, which is
surprisingly independent of the coupling strength!

Similarly, the parity is not well defined at this energy. It is just the
crossing point corresponding to doubly degenerate states. Because $f_{n}$ is
also analytic at this point, the pole curves (\ref{pole}) should also pass
through these crossing points. As demonstrated in Fig. \ref{spectra} that all
these crossing points for different $n$ (blue squares) just situate on a
horizontal line $E=-\frac{\Delta }{U}\ $ in the energy spectra. They are
usually the last crossing points for each pair of levels with positive and
negative parity and somehow hardly discerned without analytical reasonings.

Interestingly we can give a lower bound for number of states below $E=-\frac{%
\Delta }{U}$ for given $g$ by counting the  level crossing points. According to Eq. (\ref%
{gcn}), we get maximum number $n_{\max }\ $for $g_{c}^{(n)}<g$,
\begin{equation}
n_{\max }=\left[ \frac{g^{2}}{\left( 1-\frac{U^{2}}{4}\right) }-\frac{\Delta
}{U}\right] .  \label{N-num}
\end{equation}%
where the bracket [...] denotes the Gaussian step function. There are $%
n_{\max }+1$ level crossings at the same energy $-\frac{\Delta }{U}$ in the
coupling regime $[0,g]$, as shown in Fig. \ref{spectra}. Note that those
levels pass through $n_{\max }+1$ crossing points will lie below $%
E_{n}^{cross}=-\frac{\Delta }{U}$ at $g$. Then for given $g$, we find at
least $2\left( n_{\max }+1\right) $ states below $-\frac{\Delta }{U}$. In
the limit $U\rightarrow \pm 2$,$\ n_{\max }\rightarrow \infty $ by Eq. (\ref%
{N-num}). So at $U=\pm 2$, there are possibly an infinite number of levels
below or equal to $\mp \Delta /2$ for any $g$.

All other Juddian solutions for doubly-degenerate states can be figured out
in terms of the $n>0$ pole energy $E_{n}^{pole}$ (\ref{pole}). It is
required that the numerator of $f_{n}$ vanishes. For example, for $n=1$
pole,
\[
E_{1}^{pole}=\left( 1-\frac{U^{2}}{4}\right) -\frac{U\Delta }{4}-g^{2},
\]%
$\Omega _{0}$ and $\Omega _{1}$ can be obtained through Eq. (\ref{omega}),
and substitution to Eq. (\ref{recur}) yields
\[
f_{1}=\frac{2\left( 1+E-w^{2}+2gw\right) \Omega _{0}+\Delta +\left(
w^{2}-1\right) U}{n\left[ wU+2\left( g-w\right) \Omega _{1}\right] }f_{0},
\]%
It requires numerator to be zero, i.e.%
\[
2\left( 1+E_{1}^{pole}-w^{2}+2gw\right) \Omega _{0}+\Delta +\left(
w^{2}-1\right) U=0,
\]%
this is just the constrained condition. Therefore we can obtain several $g$
for $n=1$ pole curve in the energy spectra for fixed $\Delta $ and $U$.

The constrained condition becomes more complicated with larger $n$, but in
principle can be obtained. Proceeding along this line, we can predict the
values of $g$ for fixed $\Delta ,U $ in the spectra for any $n>0$. By the
way, the largest $g$ for the crossing points obtained in this way should be
the same as that in Eq. (\ref{gcn}), as stated before. These predicted
values coincide with the level crossing marked by blue filled circles and
squares, as exhibited in Fig. \ref{spectra}.

So level crossings only happen in the pole curves. We want to point out
that, except the level crossing points situating on the pole curves
described by Eqs. (\ref{pole}) and (\ref{0pole}), there are no other true
level crossings, no matter how close they are.

\subsubsection{ Exceptional solution for nondegenerate states}

\begin{figure}[tbp]
\centering
\includegraphics[width=12cm]{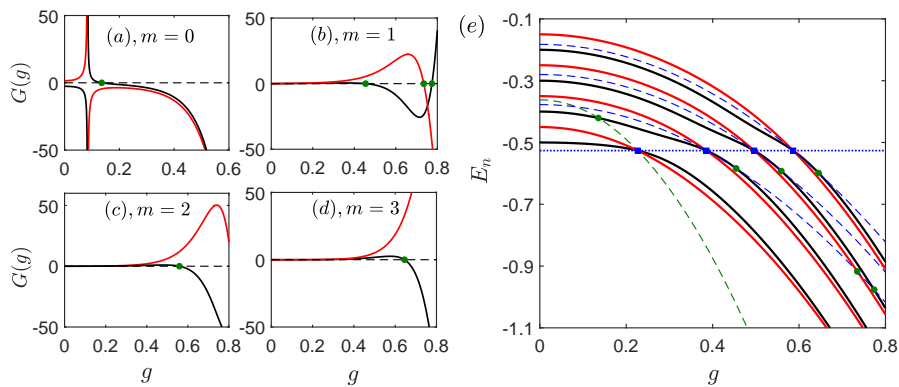}
\caption{(Color online) (a)-(d) Exceptional G-curves by Eq. (\protect\ref%
{exc_G1}) related to different $m$-th pole curves for $\Delta =1,U =1.9$.
The exceptional solutions for the nondegenerate states are indicated by
green circles. (e) Energy spectra for $\Delta =1,U =1.9$. The green dashed
line is $E_{0}^{pole}$ and the blue dashed lines are $E_{n}^{pole}$ for $%
n=1,2,3$. The Juddian solutions are indicated by blue squares. The
exceptional solutions for nondegenerate states are also denoted by green
circles. Horizontal blue dotted lines $E=-\Delta/U$ are guides to the eye.}
\label{Nondegenerate}
\end{figure}

As in the original QRM, it is possible that the $m$-th pole line can cross
the energy levels away from the level crossings, leading to exceptional
solutions for the non-degenerate state \cite{Maciejewski21}. With fixed $%
m(m\geqslant 1)$, let $f_{n<m}=0$ and $f_{n=m}=1$, the nondegenerate
exceptional $G$-function in $\Delta -g$ space is defined as \cite{Braak2015}
\begin{equation}
G_{m}^{exc}(g)=\sum_{n=m}^{\infty }\left( \Omega _{n}\pm 1\right)
f_{n}w^{n}=0  \label{exc_G1}
\end{equation}%
where the energy is limited to $E=E_{m}^{pole}$ by Eq. (\ref{pole}). The
zeros of exceptional $G$-function (\ref{exc_G1}) will give the coupling
strength.

Particularly, for the zeroth pole, c. f. Eq. (\ref{0pole}), the summation in
the $G$-function is then the same as in Eq. (\ref{G-func}). Note however
that $\Omega _{0}$ in the first term diverges if using $E_{0}^{pole}$. In
this case, we can start with $e_{0}=1$ in recurrence relation instead, so
all coefficients are well defined, including $f_{0}=0$ due to $1/\Omega
_{0}=0$. The nondegenerate exceptional $G$-function is thus
\begin{equation}
G_{0}^{exc}\left( g\right) =\sum_{n=0}^{\infty }\left( 1\pm 1/\Omega
_{n}\right) e_{n}w^{n}=0.  \label{exc_G2}
\end{equation}

We exhibit the nondegenerate $G$-function as a function of $g$ for $\Delta
=1,U=1.9$ in Fig.~\ref{Nondegenerate} (a)-(d) for $m=0,1,2,3$. The zeros
give the coupling strength where the non-degenerate exceptional solution
occurs, which are marked with green circles. To show the precise location of
these nondegenerate exceptional solutions, we also calculate the energy
spectra with the same parameters, which is displayed in Fig.~\ref%
{Nondegenerate} (e). The green circles for $m=0,1,2,3$ are also marked,
which are just the zeros exhibited in Fig.~\ref{Nondegenerate} (a)-(d).

So far, the energy spectra of the quantum Rabi-Stark model for the regular
type and two kinds of exceptional ones are completely obtained.

\section{Spectral accumulation and collapse at $U=\pm 2$}

Note that the present $G$-function (\ref{G-func}) is not valid at $U=\pm 2$,
because $w\ $in Eqs. (\ref{omega}) diverges. The spectral
phenomena  at $U=\pm 2$ should be studied in another way. We here present our analysis for $U=2$ by a new approach in detail. For
the case of $U=-2$, the extension is achieved straightforwardly by changing $%
\Delta $ into $-\Delta $.

The bosonic components of Hamiltonian can be expressed in terms of the
effective position and momentum operators of a particle of mass $m$, defined
as
\begin{equation}
x=\sqrt{\frac{1}{2m\omega }}\left( a^{\dagger }+a\right) ,p=i\sqrt{\frac{%
m\omega }{2}}\left( a^{\dagger }-a\right) ,  \label{represent}
\end{equation}%
for simplicity we can set $m\omega =1$. In terms of two eigenstates of $\sigma_z$, the Hamiltonian $H_{0}=H_{R}+H_{NL}$
in the matrix form then takes
\begin{equation}
H_0=\left(
\begin{array}{cc}
p^{2}+x^{2}-1+\frac{\Delta }{2} & g\sqrt{2}x \\
g\sqrt{2}x & -\frac{\Delta }{2}%
\end{array}%
\right) .  \label{H_U2}
\end{equation}%
Suppose the wavefunction is $\Psi =\left( \Psi _{1},\Psi _{2}\right) ^{T}$, we have two coupled Schr\"{o}dinger equations
\begin{eqnarray*}
\left( p^{2}+x^{2}-1+\frac{\Delta }{2}\right) \Psi _{1}+g\sqrt{2}x\Psi _{2}
&=&E\Psi _{1}, \\
-\frac{\Delta }{2}\Psi _{2}+g\sqrt{2}x\Psi _{1} &=&E\Psi _{2}.
\end{eqnarray*}%
Inserting $\ \Psi _{2}=\frac{g\sqrt{2}x}{E+\frac{\Delta }{2}}\Psi _{1}$ to
the first equation results in the effective one-body Hamiltonian for $\Psi
_{1},$
\[
H_{eff}\Psi _{1}=\left( E+1-\frac{\Delta }{2}\right) \Psi _{1},
\]%
where
\begin{equation}
H_{eff}=2\left( \frac{p^{2}}{2}+\frac{1}{2}\omega _{eff}^{2}\ x^{2}\right) ,
\label{H_eff}
\end{equation}%
with%
\[
\quad \omega _{eff}=\sqrt{1+\frac{2g^{2}}{\frac{\Delta }{2}+E}}.
\]

One can easily find the eigenvalues of this quantum harmonic oscillator%
\begin{equation}
E+1-\frac{\Delta }{2}=2\omega _{eff}\left( n+\frac{1}{2}\right) ,\quad
n=0,1,2,...\infty  \label{energy}
\end{equation}

To have the real harmonic frequency, $1+\frac{2g^{2}}{\frac{\Delta }{2}+E}$
should be positive, which results in $E>-\frac{\Delta }{2}$ or $E<-\frac{%
\Delta }{2}-2g^{2}$. For $E>-\frac{\Delta }{2}$, we have the equation for
the energy

\begin{equation}
\frac{\sqrt{E+\frac{\Delta }{2}}\left( E+1-\frac{\Delta }{2}\right) }{\sqrt{%
\frac{\Delta }{2}+E+2g^{2}}}=2n+1,\quad n=0,1,2,...\infty ,  \label{upper}
\end{equation}%
while for $E<-\frac{\Delta }{2}-2g^{2}$, we have another equation for the
energy

\begin{equation}
\frac{\sqrt{-\left( \frac{\Delta }{2}+E\right) }\left( E+1-\frac{\Delta }{2}%
\right) }{\sqrt{-\left( \frac{\Delta }{2}+E+2g^{2}\right) }}=2n+1,\quad
n=0,1,2,...\infty .  \label{low}
\end{equation}%
These two equations are exactly the same as Eqs. (39-41) \cite{Maciejewski2}
by Maciejewski et al. Our solution based on a harmonic oscillator is much
simpler. It must be related to the fact that Maciejewski et al. uses the
Bargmann space and transform the equations first into the so-called Birkhoff
form, apparently creating unnecessary complications. They obtain also
Hermite polynomials for the eigenfunctions in Bargmann space, but we would
say that the Hermite polynomials are the wave-functions in the ordinary
Hilbert space because the system is just a harmonic oscillator with shifted
frequency.

From Eq. (\ref{low}), we can see that an infinite number of discrete energy
levels is confined in the energy interval
\begin{equation}
\frac{\Delta }{2}-1<E<-\frac{\Delta }{2}-2g^{2},  \label{Interval}
\end{equation}%
if $g<\sqrt{\frac{1-\Delta }{2}}\label{g_c+}$. For convenience, we denote $%
E_{c}^{+}=-\Delta /2-2g^{2}$ and $g_{c}^{+}=\sqrt{\left( 1-\Delta \right) /2}
$. The effective potential becomes flat if $\ \omega _{eff}=0$, i.e. $%
E=E_{c}^{+}$.$\ $In this case, there are qubit states which turn the
potential flat \cite{Felicetti,Penna}, and the spectrum collapses, like for
a free particle. The infinite discrete energy levels in the low energy
region for $g<g_{c}^{+}$ would collapse to $E_{c}^{+}$ for $g=g_{c}^{+}$.

\begin{figure}[tbp]
\centering
\includegraphics[width=12cm]{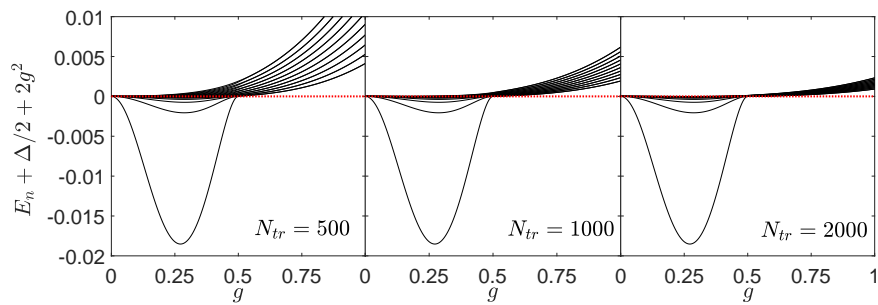}
\caption{ (Color online) The differences of the first several energy levels
and $E_{c}^{+}=-\Delta/2-2g^2$, i.e. $E_n+\Delta/2+2g^2$, as a function of $g
$ by numerical exact diagonalizations with the truncation number $N_{tr}=500$
(left),$1000$ (middle), and $2000$ (right) for $U=2$, $\Delta =0.5$, and
accordingly $g_{c}^{+}=0.5$. The red horizontal line corresponds to the
energy value $E_{c}^{+}$.}
\label{wholecoupling}
\end{figure}

For $g>g_{c}^{+}$, from Eqs. (\ref{low}) and (\ref{Interval}), we know that
no real solutions exist in this case. Then we have to resort to numerics. In
Fig. ~\ref{wholecoupling}, we exhibit the first several energy levels for $%
U=2,\Delta =0.5$ with different truncation of the Fock space by numerical
exact diagonalziation. $g_{c}^{+}=0.5$ in this case. All energies for $%
g>g_{c}^{+}$ become closer to $E_{c}^{+}$ monotonously with increasing
truncated photonic number $N_{tr}$, although the convergence is hardly
achieved by numerics. It is observed that $E_{c}^{+}$ is a lower bound in
the regime of $g>g_{c}^{+}$. In the two-photon QRM ~\cite{Felicetti,duan},
it can be easily checked that the energy in numerical diagonalization has no
a lower bound when coupling strength is larger than the half of cavity
frequency. Although both models have a common feature of spectral collapse
at a critical coupling, they display essentially different behaviour above
the critical coupling.

While for $g<g_{c}^{+}$, one can see from the left-hand-side of each plot in
Fig. ~\ref{wholecoupling} that the converging energies for low excited
states in the present model are easily obtained numerically, which can be
also confirmed by the solution to Eq. (\ref{low}). However, it is extremely
difficult to obtain the converging energy level by direct exact
diagonalizations when energy approaches to $E_{c}^{+}$. Close to $E_{c}^{+}$%
, there is a quasi-continuum of states with an infinite number of discrete
states.

We can analyze the average photonic number $N$ in each eigenstates.
According to the effective harmonic oscillator (\ref{H_eff}), we have the
wavefunction in the $n$-th energy level
\begin{equation}
\Psi _{n}\varpropto \left(
\begin{array}{c}
1 \\
\frac{g\sqrt{2}x}{E+\frac{\Delta }{2}}%
\end{array}%
\right) H_{n}\left( \omega _{eff}x\right),  \label{wave}
\end{equation}%
where $H_{n}$ is the Hermite polynomial of degree $n$. Then we can calculate
$N$ straightforwardly, which is however very tedious and not shown here.
When $E\rightarrow E_{c}^{+},$ $N$ is approximately equal to%
\begin{equation}
N\approx g^{2}+\frac{3g^{2}\left( \Delta +2g^{2}-1\right) }{4\left(
E_{n}-E_{c}^{+}\right) }+\frac{3}{8\left( 1-\Delta -2g^{2}\right) },
\label{Number}
\end{equation}%
where the use has been made of Eqs. (\ref{represent}) and (\ref{energy}).

One can find from Eq. (\ref{Number}) that $N$ diverges if energy level
approaches to $E_{c}^{+}$. So it is almost impossible to use numerical exact
diagonalziation to calculate correctly the energy level if very close to $%
E_{c}^{+}$. If the numerical truncation of the Fock space is below $N$, the
results depend naturally on the truncation. This is a case where only an
analytical treatment can give the correct answer.

The high energy levels for $E>-\Delta /2$ for arbitrary coupling $g\ $can be
easily obtained by Eq. (\ref{upper}) analytically. Our hypothesis for the
exotic energy distribution for $E<-\Delta /2$ at $U=2$ is the following. For
$g<g_{c}^{+}$, by Eq. (\ref{low}), we know that infinitely many discrete
levels lie below accumulation point $E_{c}^{+}$. Close to $E_{c}^{+}$, there
is a quasicontinuum of states. All states are normalizable. They collapse to
$E_{c}^{+}$ right at $g_{c}^{+}$. When $g>g_{c}^{+}$ all these energy levels
could only stay in energy interval $E_{c}^{+}\leq E<-\frac{\Delta }{2}$, but
absolutely cannot be given by Eq. (\ref{low}). The corresponding states
should be unnormalizable. In this energy interval, it is unclear whether
there is a continuum of non-noramlizable states, or this region is empty,
which remains an open question. This issue obviously could not be addressed
by any numerics and the above analytical theory in the framework of a
harmonic oscillator, therefore other rigorous study should be called for.

\section{ Conclusion}

In this work, we have derived the $G$-function for the quantum Rabi-Stark
model in a compact way by using the BOA. Zeros of the $G$-function determine
the regular spectrum. Two kinds of exceptional solutions are clarified and
demonstrated. For the Juddian-type solution, the true level crossing occurs
at the doubly degenerate states with both parities, which exclude the
previous "crossing" from the same parity. The first-order phase transition
is detected analytically by the pole structure of $G$-functions. The
critical coupling strength of the phase transitions is obtained
analytically. The exotic energy spectra at $U=\pm 2$ are analyzed within an
effective quantum harmonic oscillator. Previous energy spectra by very
complicated and cumbersome derivation can be very easily reproduced.
Moreover, the energy spectral collapse can be attributed by the flat
quadratic potential. Below the collapse critical coupling, there are
infinite discrete energy levels below the collapse energy.

Both the first-order quantum phase transition and the spectral collapse can
occur in the present model, and are lacking in the linear QRM. The spectral
collapse also occurs in the two-photon QRM with another kind of nonlinear
coupling, but the first-order quantum phase transition is absent. Spectral
collapse does not occur in the anisotropic QRM where the first-order phase
transitions can be induced by the anisotropy with respect to the
rotating-wave and non-rotating-wave coupling strengths. It follows that the
Stark-like nonlinear coupling between atom and cavity is of fundamental
importance. We believe that the quantum Rabi-Stark model would exhibit
various fundamental phenomena found in the various QRMs, and could even go
beyond. The spectral collapse and the discrete levels below the collapse
energy might be qualitatively understood in the polaron picture by the
tunneling induced potential well \cite{conglei}. Due to the parity symmetry,
the second-order phase transition in the present model should also occur in
the limit $\Delta /\omega \rightarrow \infty $, like that in the linear QRM ~%
\cite{plenio,hgluo,Shen, Ashhab1}. We speculate that the present model would
possibly experience true superradiance transition in the single-atom model
at moderate frequency ration $\Delta /\omega$. Other peculiarities and novel
properties in quantum Rabi-Stark model are also worthy of further
explorations. The well understanding of the closed system will lay the solid
foundation for further treatment of the open quantum system \cite{Dimer}.

\ack{ We acknowledge useful discussions with Daniel Braak, Hans-Peter Eckle, and
Stefan Kirchner. This work is
supported by the National Science Foundation of China (Nos. 11674285,
11834005), the National Key Research and Development Program of China (No.
2017YFA0303002).}

\end{document}